\documentclass[]{spie}  

\usepackage{amsmath,amsfonts,amssymb}
\usepackage{graphicx}
\usepackage[colorlinks=true, allcolors=blue]{hyperref}
\usepackage{graphicx}
\usepackage{lipsum}
\usepackage{booktabs}
\usepackage{amsfonts}
\usepackage{booktabs}
\usepackage{siunitx}

\title{Transformer-Based Local Feature Matching for
Multimodal Image Registration}

\author[a,b]{Remi Delaunay}
\author[a,b]{Ruisi Zhang}
\author[a,b,c,d]{Filipe C. Pedrosa}
\author[c,d]{Navid Feizi}
\author[b,e]{Dianne Sacco}
\author[c,d]{Rajni Patel}
\author[a,b]{Jayender Jagadeesan}
\affil[a]{Brigham and Women's Hospital, MA, USA.}
\affil[b]{Harvard Medical School, MA, USA.}
\affil[c]{Western University, ON, Canada.}
\affil[d]{Canadian Surgical Technologies and Advanced Robotics, ON, Canada.}
\affil[e]{Massachusetts General Hospital, MA, USA.}

\authorinfo{Further author information: (Send correspondence to Remi Delaunay)\\E-mail: delaunay.remi@gmail.com}

\begin{document} 
\thispagestyle{empty}
\newpage
\thispagestyle{empty} 
\vspace*{\fill} 
{\Large
This manuscript has been accepted to SPIE Medical Imaging 2024. Please use the following reference when citing the manuscript:
\vspace{0.5cm}

Remi Delaunay, Ruisi Zhang, Filipe C. Pedrosa, Navid Feizi, Dianne Sacco, Rajni V. Patel, Jayender Jagadeesan, "Transformer-based local feature matching for multimodal image registration," Proc. SPIE 12926, Medical Imaging 2024: Image Processing, 129260I; \textbf{https://doi.org/10.1117/12.3005591}}
\vspace*{\fill}

\newpage

\maketitle

\begin{abstract}
Ultrasound imaging is a cost-effective and radiation-free modality for visualizing anatomical structures in real-time, making it ideal for guiding surgical interventions. However, its limited field-of-view, speckle noise, and imaging artifacts make it difficult to interpret the images for inexperienced users. In this paper, we propose a new 2D ultrasound to 3D CT registration method to improve surgical guidance during ultrasound-guided interventions.  Our approach adopts a dense feature matching method called LoFTR to our multimodal registration problem. We learn to predict dense coarse-to-fine correspondences using a Transformer-based architecture to estimate a robust rigid transformation between a 2D ultrasound frame and a CT scan. Additionally, a fully differentiable pose estimation method is introduced, optimizing LoFTR on pose estimation error during training. Experiments conducted on a multimodal dataset of \textit{ex vivo} porcine kidneys demonstrate the method's promising results for intraoperative, trackerless ultrasound pose estimation. By mapping 2D ultrasound frames into the 3D CT volume space, the method provides intraoperative guidance, potentially improving surgical workflows and image interpretation.
\end{abstract}


\section{INTRODUCTION}
\label{sec:intro}  

Ultrasound imaging is a cost-effective and radiation-free modality that allows to visualize anatomical structures and monitor changes in real-time during procedures, making it ideal for guiding surgical interventions. However, its limited field-of-view, speckle noise appearance, and imaging artifacts (e.g., shadowing from bones and stones) makes it difficult to navigate with for inexperienced users. Therefore, ultrasound imaging is typically used in conjunction with a pre-operative scan (e.g., CT or MRI) to provide a more comprehensive view of the anatomy and guide the sonographer during acquisition.

This work focuses on the registration of 2D ultrasound frames with a CT volume. While different methods for multimodal image registration have been proposed, they usually fail if no initial alignment is provided \cite{wein2013global,san2009towards,hering2022learn2reg}. This is mainly due to the difference in image appearance and contrast, which makes it challenging to find corresponding features between images. In ultrasound-guided interventions, this initial pose is often obtained by tracking the position of the ultrasound probe during acquisition using an external optical tracker \cite{shapey2021integrated}, and registering the tracker coordinates with the diagnostic CT frame. However, this technique is impractical for use in many clinical settings since it requires additional hardware and calibration steps that are complex to set up, and may slow down the surgical workflow. Another promising approach consists of matching predicted segmentation maps to initialize the registration process \cite{montana2021vessel}.

Learning-based methods are now at the forefront of emerging computer vision approaches for the detection of matching descriptors in images \cite{detone2018superpoint,sarlin2020superglue}. These methods typically use a convolutional neural network (CNN) to learn complex feature representations, and have been shown to outperform traditional handcrafted feature-based methods such as SIFT~\cite{lowe1999object} or SURF~\cite{bay2008speeded}. A method called ``Detector-Free Local Feature Matching with Transformers" (LoFTR) has been recently introduced for local image feature matching, using a transformer-based architecture to establish semi-dense matches at a coarse-level, and refine them at a fine-level \cite{sun2021loftr}. Inspired by LoFTR, Markova et al. proposed a U-Net architecture that predicts keypoint descriptors, from which they estimate a global transformation for ultrasound-to-MR registration \cite{markova2022global}.

In this paper, we propose a modified version of the LoFTR method to overcome the challenges of \textbf{trackerless} ultrasound pose estimation in image-guided surgery. By adapting the Transformer-based approach to multimodal and multidimensional image data, we can efficiently predict coarse-to-fine correspondences between a 2D ultrasound frame and a preoperative 3D CT scan, to estimate a rigid transformation. Furthermore, we propose a fully differentiable pose estimation method that allows direct optimization of LoFTR on pose prediction error during training. We demonstrate the performance of the method on a multimodal dataset of \textit{ex vivo} porcine kidneys and show promising results for intraoperative, image-based ultrasound pose estimation. Fully automatic, multimodal, trackerless image fusion could potentially improve the surgical workflow and image interpretation for ultrasound-guided interventions.
\section{METHODS}

\begin{figure}
\centering
\includegraphics[width=\textwidth]{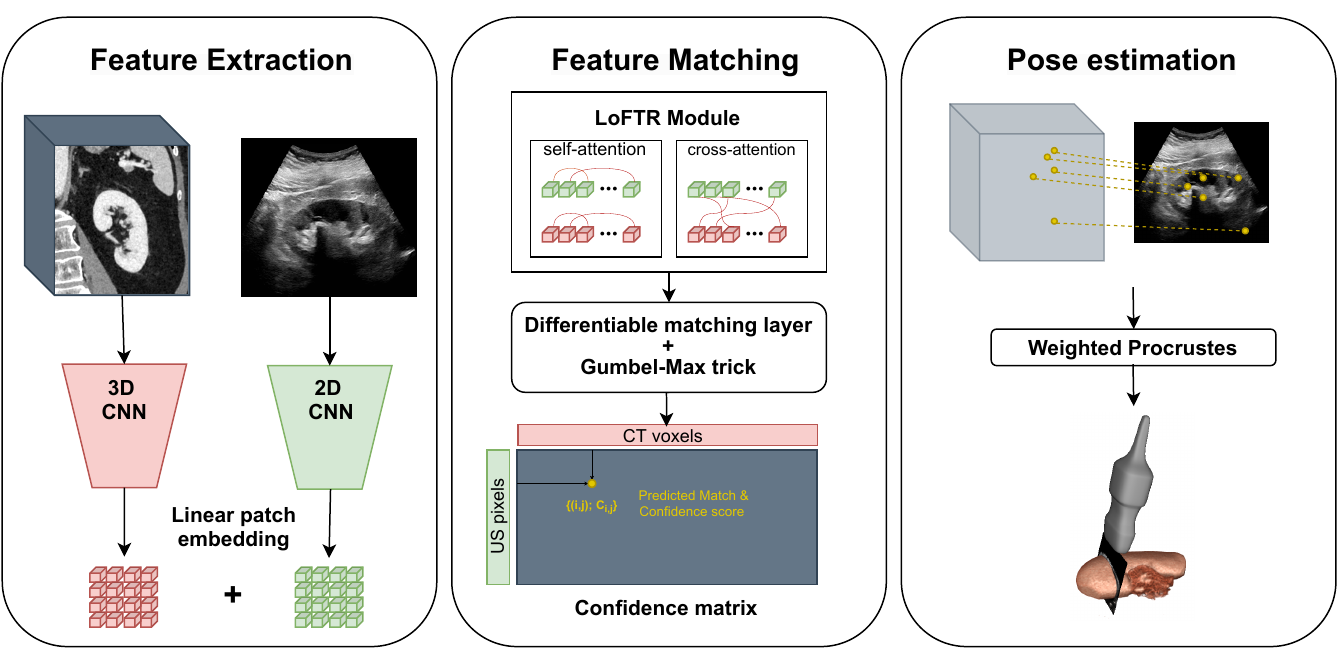}
\caption{Overview of the proposed method. A CNN is first employed to extract both coarse and fine-level feature maps from the input image pairs, which are then converted into linear patch embedding to be processed by the LoFTR module. Then, a differentiable matching layer is utilized to match the transformed features, resulting in a confidence matrix $P_{M_c}$. Finally, the predicted matches and confidence scores are used to predict a rigid transformation with the Differentiable Weighted Procrustes (DWP) method.} \label{fig1}
\end{figure}

The objective of this work is to find a global transformation that maps a 2D ultrasound frame into the space of a preoperative 3D CT volume, which can then be used for intraoperative guidance. An overview of the proposed approach is shown in Figure~\ref{fig1}. We adapted the LoFTR method, which makes use of a hybrid CNN-Transformer architecture to extract and match local features from the multimodal input image pairs \cite{sun2021loftr}. Finally, the matched feature points are used to retrieve a rigid pose by using our new differentiable pose estimation strategy. The rest of this section provides a summary of the different components of our method, and highlights the changes made to adapt it to our registration problem.

\subsection{Feature extraction}
We use a standard ResNet-18 backbone combined with a Feature Pyramidal Network (FPN) to extract the local features from our data \cite{lin2017feature}. The FPN is a well-known network architecture which efficiently extracts feature maps with both low-level and high-level information from the image. Multimodal coarse-level features, $\tilde{F}^{US}$ and $\tilde{F}^{CT}$, and fine-level features, $\hat{F}^{US}$ and $\hat{F}^{CT}$ are extracted at 1/8th and 1/4th of their respective original image dimensions. Two separate models are employed to account for the difference in dimensions of our input image pairs: a 2D network for the ultrasound frame and a 3D one for the CT volume. Before feeding them into the LoFTR module, the feature maps are flattened into 1D vectors and summed with 2D/3D positional encoding to inject information about the position of each element into the input sequence. The flattened features can be thought of as a sequence of image patches, where each patch corresponds to a position in the 1D vector.
\subsection{Feature matching}

We use the LoFTR method \cite{sun2021loftr} to produce cross-modality feature representations that promote feature matching on the ultrasound and CT images. The LoFTR architecture is based on a memory-efficient version of the Transformer network, known as linear Transformer \cite{katharopoulos2020transformers}. As opposed to convolutional networks that rely on local receptive fields, Transformer networks have been shown to better capture the global structure of images in different medical imaging domains, including registration. The LoFTR module further enhances the learning of spatial relationships between the two feature inputs, by sequentially attaching self-attention and cross-attention layers $N_f$ times to generate more informative feature representations before matching.
The matching probability $P_{M_c}(i,j)$ of the transformed coarse-level features $\tilde{F}^{US}_{tr}$ and $\tilde{F}^{CT}_{tr}$ is determined by applying the differentiable dual-softmax operator over the score matrix $S(i,j) = \langle \tilde{F}^{US}_{tr}, \tilde{F}^{CT}_{tr} \rangle$:

\begin{equation}
    P_{M_c}(i,j) = \mathrm{Softmax}(S(i,\cdot))_j \cdot \mathrm{Softmax}(S(\cdot,j))_i
\end{equation}

\subsection{Differentiable pose estimation}

In the original LoFTR approach, matches between the two feature sets are extracted by applying a hard threshold on the confidence matrix values. However, the discrete sampling of matches introduces a non-differentiability issue, breaking the backpropagation process. Consequently, the inability to predict a rigid transformation using a differentiable solver hinders the inclusion of pose estimation error during training. Thus, the coarse-level supervision in LoFTR relies solely on the predicted confidence matrix $P_{M_c}$.

To address this limitation, we propose a modified approach that overcomes the challenges of discrete sampling. We employ the Gumbel-Softmax reparameterization trick, which allows us to sample matches from the confidence matrix in a continuous and differentiable manner \cite{jang2016categorical}. Additionally, we incorporate the Differentiable Weighted Procrustes estimator (DWP), a technique used to find an optimal rigid transformation between two sets of points while considering individual weights \cite{choy2020deep}. By passing gradients through the sampled confidence scores (weights) associated with each set of matches, we achieve end-to-end training. As a result, our method effectively incorporates pose estimation error during training, thereby enhancing the coarse-level supervision and contributing to improved performance in the feature matching process.

\section{EXPERIMENTS}
\label{sec:title}
\subsection{Datasets}

Our multimodal dataset consists of images obtained from 10 \textit{ex vivo} freshly excised porcine kidneys embedded in medium density gel wax (Penreco, PA, USA). A 3D CT scan with 0.6 mm slice thickness, and 12 ultrasound sweeps of approximately 1700 images were acquired for each \textit{ex vivo} kidney phantom, resulting in a total of \num{201528} ultrasound frames. The ultrasound sweeps were acquired from the top of the phantom with varied orientations and directions, to enrich the dataset. The 2D ultrasound frames were collected with a BK ProFocus ultrasound system (BK Medical, MA, USA), and their poses in world coordinates were obtained by attaching a custom-made passive marker to the ultrasound probe (8820e curved-array), and tracking its position with the Polaris Vicra optical tracking system (NDI, ON, Canada). Ground truth pose between the ultrasound sweeps and CT scan from the kidney phantom dataset was obtained by placing CT-compatible fiducials on each phantom, and registering the CT scan with the physical phantom. The pose was then refined by using an optimization-based method and manual verification. Ultrasound acquisition, calibration and registration was performed using the ultrasound module from the ImFusion Suite 2.44.1 (ImFusion GmbH, Munich, Germany) software. 


\subsection{Results}

The baseline method for comparison with our modified version of LoFTR is the ``BRAINSFit" general registration framework based on ITK \cite{johnson2007brainsfit}. We used the default parameters provided for volumetric multimodal registration by mutual information optimization. We compared the results of BRAINSFit by first aligning the centers of the two volumes (Baseline) and then by using the transform predicted by our approach (LoFTR-DWP + Baseline). We also compared the results of the original LoFTR rigid transform estimation (LoFTR-RANSAC), which consists in using the RANSAC estimator in post-processing, to LoFTR-DWP to quantify the contribution of our pose error cost function during training. 

\begin{figure}
\centering
\includegraphics[width=\textwidth]{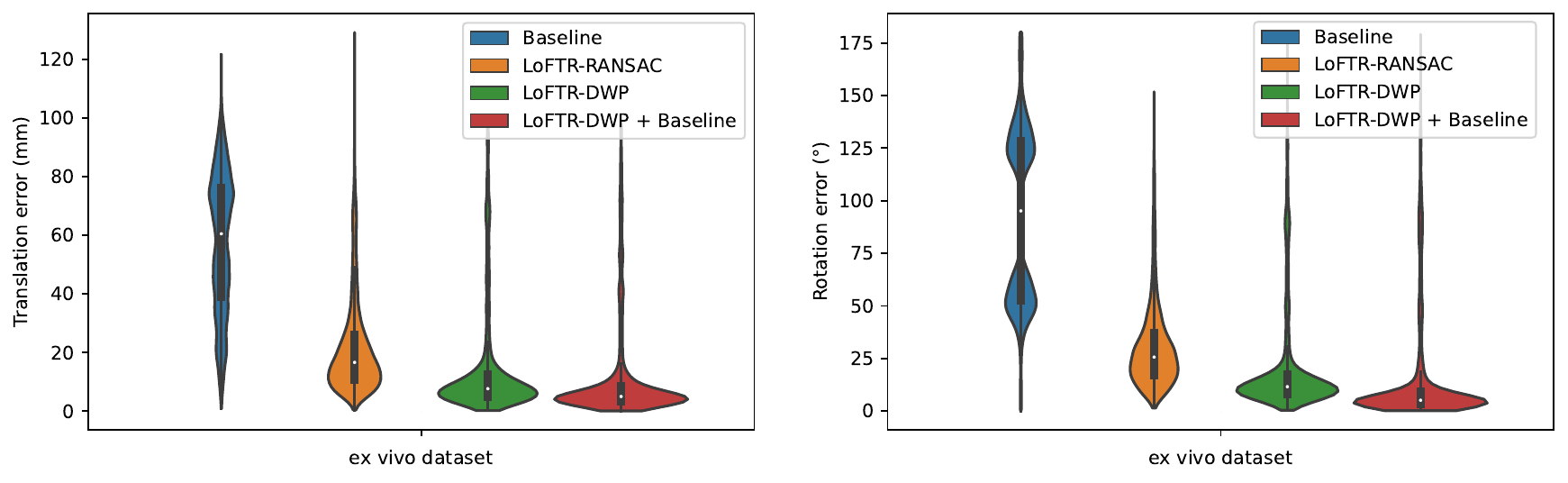}
\caption{Pose prediction error distribution as a violin plot} \label{fig2}
\end{figure}

We quantified the registration accuracy in terms of mean, standard deviation and median rotation error (in deg) and translation error (in mm) between the estimated and ground truth ultrasound pose (see  \ref{tab:1}). In addition, we included the percentage of cases that are below the accepted accuracy threshold for surgical navigation, i.e., rotation $< 5^\circ$ and translation $< 5$ mm. Table \ref{fig1} also contains the percentage of cases that were considered successful in terms of initial pose estimation (i.e., rotation $< 15^\circ$ and translation $< 15 $ mm) for optimization-based registration \cite{wein2013global}. Finally, we show in Figure \ref{fig2} a violin plot of the distribution of pose errors for a more comprehensive interpretation of the registration performance of each compared methods.

\begin{table}[!ht]
\centering
\caption{Comparison of pose prediction error.}
\resizebox{\textwidth}{!}{
\begin{tabular}{lcccccc} \toprule
    & \textbf{Baseline} & \textbf{LoFTR-RANSAC} & \textbf{LoFTR-DWP} & \textbf{LoFTR-DWP + Baseline} \\ \midrule
    Rot. (deg) & 91.77 $\pm$ 41.34 & 29.83 $\pm$ 19.30 & 20.23 $\pm$ 24.68 & 15.09 $\pm$ 26.07 \\
    & (95.43) & (25.59) & (11.51) & (5.06) \\
    Trans. (mm) & 106.78 $\pm$ 19.13 & 21.56 $\pm$ 16.48 & 14.14 $\pm$ 18.24 & 11.53 $\pm$ 17.65 \\
    & (112.17) & (11.51) & (7.62) & (4.95) \\ \midrule
    $15^\circ$ \& 15 mm & 0\% & 16\% & 67\% & 81\% \\
    $5^\circ$ \& 5 mm & 0\% & 1\% & 6\% & 39\% \\ \bottomrule
\end{tabular}
}
\label{tab:1}
\end{table}

\section{Discussion}

In this paper, we adapted a transformer-based method to multimodal and multidimensional image data for ultrasound pose estimation in image-guided surgery. Our approach predicts dense correspondences between a given 2D ultrasound frame and a 3D preoperative CT, which are then used to estimate a rigid transformation. The Baseline method shows significantly higher pose prediction errors compared to the proposed method. Optimization-based methods rely on the optimization of similarity metrics, which often get stuck in local minima when no meaningful initial alignment is provided. The LoFTR-DWP method shows a substantial improvement over both Baseline and LoFTR-RANSAC in terms of pose prediction errors, with 67 \% of predicted poses having a successful initial pose estimation. This suggests that using our differentiable pose estimation module during training has effectively contributed to reducing the pose prediction errors.  When provided with the predicted pose from LoFTR-DWP as an initial transform, the baseline method further improves the pose prediction accuracy, with $\SI{5.06}{\degree}$ rotation and $4.95 mm$ translation median pose error.

Finally, this work demonstrates the versatility of computer vision feature matching techniques to other image domains. By fully automating the multimodal image fusion, we believe our approach has the potential to improve the surgical workflow in ultrasound-guided procedures.


\bibliography{mybibliography} 
\bibliographystyle{spiebib} 

\end{document}